\documentclass[final,5p,times,twocolumn,letterpaper]{elsarticle}
\biboptions{sort&compress}
\usepackage{slashed}
\usepackage{eucal} 
\usepackage{bm}
\usepackage[hypertexnames=false, colorlinks=true, citecolor=blue, urlcolor=blue, linkcolor=black]{hyperref}
\usepackage{amsmath,amssymb,amsfonts,latexsym}
\newcommand{\Sthree}{S_{\hspace{-.3ex}3}}
\newcommand{\Sthreep}{{S'}_{\hspace{-.8ex}3}}
\allowdisplaybreaks
\begin{document}
\begin{frontmatter}
\author[1]{June-Young Kim}
\ead{jun-young.kim@inha.ac.kr}
\author[2]{Ho-Yeon Won}
\ead{hoyeon.won@polytechnique.edu}
\author[3]{Christian Weiss}
\ead{weiss@jlab.org}
\affiliation[1]{organization={Department of Physics, Inha University},
postcode={402-751},
postcodesep={},
city={Incheon},
country={Republic of Korea}}
\affiliation[2]{organization={CPHT, CNRS, \'Ecole polytechnique, Institut Polytechnique de Paris},
postcode={},
postcodesep={},
city={Palaiseau},
country={France}}
\affiliation[3]{organization={Theory Center, Jefferson Lab},
city={Newport News},
postcode={Virginia 23606},
postcodesep={,},
country={USA}}
\title{Quark orbital angular momentum as a chiral magnetic effect}
\begin{abstract}
The flavor-nonsinglet ($u - d$) quark angular momentum (AM) in the proton is computed based on
the effective spin-flavor dynamics emerging from chiral symmetry breaking by QCD instantons.
The QCD AM operators are converted to effective spin-flavor operators expressing
instanton-induced chiral interactions. A large negative orbital AM $L_{u - d}$
arises as a ``chiral magnetic effect'' of the interaction of the quarks
with the chiral mean field in the proton in the large-$N_c$ limit. It cancels part of the large
positive spin AM $S_{u-d}$ and reduces the total AM $J_{u-d} = L_{u-d} + S_{u-d}$,
in agreement with lattice QCD calculations.
\end{abstract}
\begin{keyword}
Nucleon spin decomposition \sep instantons \sep chiral symmetry breaking \sep chiral magnetic effect
\end{keyword}
\end{frontmatter}
\section{Introduction}
Explaining the origin of the nucleon spin is a central goal of hadronic physics.
Basic questions are how the spin is composed from the QCD quark and gluon fields,
and from the spin and orbital angular momentum (AM) of the fields \cite{Leader:2013jra}.
The operators measuring the various contributions are defined in terms of the quark
and gluon parts of the QCD energy-momentum tensor (EMT).
Of particular interest is the $u - d$ flavor-nonsinglet quark AM.
The flavor-nonsinglet spin AM $S_{u-d}$ is known to be large because of its connection
with the isovector axial coupling $g_A = 2S_{u-d}$; the total AM $J_{u-d}$ is
found to be substantially smaller in lattice QCD calculations \cite{Alexandrou:2020sml,Wang:2021vqy}.
This indicates a large negative flavor-nonsinglet orbital AM $L_{u-d}$ and raises the question
of its dynamical origin. 

The effective dynamics at the hadronic scale is determined by the spontaneous breaking of
chiral symmetry in the QCD vacuum. It is caused by the presence of topologically
charged gauge fields with fermionic zero modes, which induce chirality-changing
interactions among the quarks \cite{Schafer:1996wv,Diakonov:2002fq}.
It converts the original color dynamics into an effective spin-flavor dynamics, which
gives rise to large flavor-nonsinglet structures such as $g_A$ and the nucleon
magnetic moments. One should expect that chiral symmetry breaking
plays a major role also in the flavor-nonsinglet quark AM.

In this letter we study the flavor-nonsinglet quark AM on the basis of the effective
spin-flavor dynamics emerging from dynamical chiral symmetry breaking in QCD. We aim to answer two questions:
How does chiral symmetry breaking generate a potentially large orbital AM $L_{u-d}$?
What is the decomposition of $J_{u-d}$ in $S_{u-d}$ and $L_{u-d}$?

As theoretical framework we employ: (i) The instanton vacuum \cite{Schafer:1996wv,Diakonov:2002fq},
an effective description of the topologically charged gauge fields abstracted from lattice QCD simulations using
cooling or gradient flow methods \cite{Chu:1994vi,Bonati:2014tqa,Alexandrou:2017hqw,Athenodorou:2018jwu}.
It enables explicit derivation of chiral symmetry breaking and the effective spin-flavor dynamics
\cite{Diakonov:1985eg}. It also provides the effective operators representing QCD operators
in the chirally broken vacuum (at the scale $\bar\rho \approx$ 0.3 fm),
which express the spin-flavor interactions induced by the instantons \cite{Diakonov:1995qy,Balla:1997hf}.
The effective dynamics and effective operators are constructed
in a systematic expansion in the instanton density, which preserves operator relations resulting from
the QCD equations of motion. (ii) The mean-field picture of the nucleon in the large-$N_c$ limit,
based on the effective chiral dynamics \cite{Diakonov:1987ty}.
It permits calculation of the nucleon matrix elements of the effective operators in a
systematic $1/N_c$ expansion. It also provides a mechanical interpretation of the matrix elements
in terms of the quark interactions with the chiral mean field.

The present study is made possible by recent advances in understanding the instanton contributions to the
twist-3 QCD EMT \cite{Kim:2023pll}. The methods have been used to study the spin-orbit correlations of
quarks \cite{Kim:2024cbq}. Here use them to analyze the flavor-nonsinglet quark AM in the nucleon.

\section{QCD angular momentum}
The contribution from quarks of flavor $f$ to the QCD EMT is measured by the operator
\begin{align}
T_f^{\mu \nu}(x) \equiv \bar{\psi}_f (x) \gamma^{\mu} i \overleftrightarrow{\nabla}^{\nu} \psi_f (x) ,
\hspace{.2em}
\overleftrightarrow{\nabla}^{\nu} \equiv
\tfrac{1}{2} (\overrightarrow{\partial} - \overleftarrow{\partial})^\nu
- i g A^{\nu}(x),
\label{emt_f_qcd}
\end{align}
where $\bar{\psi}_f, \psi_f(x)$ are the quark fields and $\overleftrightarrow{\nabla}^{\nu}$
is the covariant derivative with the gauge potential.
The total EMT is given by the sum of the quark and gluon contributions (see e.g.\ Ref.\cite{Lorce:2018egm}),
\begin{align}
T^{\mu\nu} = {\textstyle\sum_f} T_f^{\mu \nu} + T_g^{\mu \nu},
\label{emt_tot}
\end{align}
and is a conserved current. The individual flavor contributions Eq.~(\ref{emt_f_qcd}) are not
conserved currents. The rank-2 tensor operator Eq.~(\ref{emt_f_qcd}) is nonsymmetric and
contains a symmetric and an antisymmetric part,
$T^{\mu\nu} = \tfrac{1}{2}(T^{\{\mu\nu\}} + T^{[\mu\nu ]})$, with
$T^{\{\mu\nu\}}, T^{[\mu\nu ]} \equiv T^{\mu\nu} \pm T^{\nu\mu}$.
The antisymmetric part can be expressed as
the total derivative of the quark axial current operator,
\begin{align}
\tfrac{1}{2} 
T_f^{ [\mu \nu ]}(x) = -\tfrac{1}{4} \epsilon^{\mu \nu \alpha \beta}
\partial_{\alpha} [\bar{\psi}_f (x) \gamma_{\beta} \gamma_{5} \psi_f (x)];
\label{eom_qcd}
\end{align}
the relation follows from the QCD equations of motion
($\gamma_5 \equiv -i \gamma^{0} \gamma^{1} \gamma^{2} \gamma^{3}$,
$\epsilon^{0123}=+1$). As such the tensor operator describes both the spin and the
orbital AM of the quark fields \cite{Lorce:2017wkb,Lorce:2018egm}.

The matrix element of the tensor operator Eq.~(\ref{emt_f_qcd}) between nucleon states
with 4-momenta $p$ and $p'$ and spin quantum numbers $\lambda$ and $\lambda'$
is parametrized as \cite{Lorce:2017wkb,Lorce:2018egm}
\begin{align}
&\langle N' | T_f^{ \mu \nu }(0) | N \rangle = \bar{u}' \bigg{[}
\frac{P^{\mu} P^{\nu}}{M_{N}}A_f(t)
+ \frac{\Delta^{\mu} \Delta^{\nu} - \Delta^{2} g^{\mu \nu}}{4M_{N}} D_f(t) 
\cr
&+ \frac{P^{\{\mu} i \sigma^{\nu\} \lambda} \Delta_{\lambda}}{2 M_{N}}J_f(t)
- \frac{P^{[\mu} i \sigma^{\nu] \lambda} \Delta_{\lambda}}{2 M_{N}} S_f(t)
+ M_{N} g^{\mu \nu} \bar{c}_f (t)
\bigg{]} u,
\label{parametrization}
\end{align}
where $u, \bar u' \equiv u(p,\lambda), \bar u(p',\lambda')$
are the bispinor wave functions normalized as $\bar{u} u = 2M_{N}$;
$M_{N}$ is the nucleon mass, $P\equiv (p+p')/2$, and $\Delta \equiv p' - p$.
The form factors depend on the invariant momentum transfer $t \equiv \Delta^2$
and characterize the internal structure of the nucleon; their interpretation has been
discussed extensively in the literature. Here we focus on $J_f(t)$ and $S_f(t)$
connected with the quark AM in the nucleon (the spin-1 multipoles of the matrix element).
Due to the equation-of-motion relation Eq.~(\ref{eom_qcd}), $S_f(t)$
can be identified with the nucleon axial form factor for quark flavor $f$,
\begin{align}
S_f (t) = \tfrac{1}{2} G_{A, f}(t).
\end{align}

The AM interpretation of the form factors can be developed in a 3D formulation
in the Breit frame \cite{Polyakov:2002yz}, or in a 2D light-front formulation in the Drell-Yan-West
frame \cite{Adhikari:2016dir,Lorce:2017wkb,Granados:2019zjw};
the two formulations can be related to each other \cite{Lorce:2018egm}.
Here we adopt the 3D formulation, which is natural for the analysis in the large-$N_{c}$ limit
of QCD. The nucleon spin states are canonical and labeled by the spin projection on the 3-axis,
$\Sthree$ and $\Sthreep$. 
The spatial distribution of the EMT ($0k$ component) is defined as the 3D Fourier transform of the
transition matrix element,
\begin{align}
\langle T^{0k}_f \rangle(\bm{x}) \equiv \int \frac{d^{3} \Delta}{(2\pi)^{3}}
e^{-i \bm{x}\cdot\bm{\Delta}} \frac{\langle N' | T^{0k}_f(0) | N \rangle}{2P^{0}}.
\end{align}
The orbital AM along the 3-direction is then defined as the rotation of the nonsymmetric tensor
(for $\Sthreep = \Sthree = +\tfrac{1}{2}$),
\begin{align}
L_f &\equiv \int d^{3} x \, \epsilon^{3jk} x^{j} \langle T^{0k}_f \rangle(\bm{x})
\nonumber \\
&= - i \epsilon^{3jk} \frac{\partial}{\partial \Delta^{j}}
\left[\frac{\langle N' | T^{0k}_f | N \rangle}{2P^{0}} \right]_{\bm{\Delta} = 0}
\hspace{-1em}
= J_f(0) - S_f(0).
\hspace{1em}
\label{L_qcd}
\end{align}
The spin AM is defined as the rotation of the negative of the
antisymmetric part of the tensor,
\begin{align}
S_f &\equiv  -\tfrac{1}{2} \int d^{3} x \, \epsilon^{3jk} x^{j}   \langle T^{[0k]}_f \rangle(\bm{x})
= S_f(0) = \tfrac{1}{2} G_{A, f}(0).
\label{S_qcd}
\end{align}
The total AM is defined through the symmetric part of the tensor (or Belinfante-Rosenfeld-type EMT),
\begin{align}
J_f &\equiv  \tfrac{1}{2} \int d^{3} x \, \epsilon^{3jk} x^{j}   \langle T^{\{0k\}}_f \rangle(\bm{x})
= J_f(0).
\label{J_qcd}
\end{align}
The definitions obviously satisfy
\begin{align}
J_f = S_f + L_f,
\label{sumrule_qcd}
\end{align}
which can be interpreted as the decomposition of the total quark AM of flavor $f$
into orbital and spin AM.
Note that $J_f, S_f$ and $L_f$ are unambiguously  defined in terms of the invariant form factors of
Eq.~(\ref{parametrization}), and their values do not depend on the interpretation in the Breit frame.

\section{Chiral symmetry breaking}
The basics of chiral symmetry breaking and the QCD instanton vacuum are described in
Refs.~\cite{Schafer:1996wv,Diakonov:2002fq}. The QCD vacuum at low resolution scales is populated
by instantons and antiinstantons, strong nonperturbative gauge fields which carry topological charge $\pm 1$ and
represent tunneling trajectories between topological sectors of the gauge theory. They induce fermionic
zero modes with fixed chirality, which cause the breaking of chiral symmetry of light quarks.
The instanton density in 4D Euclidean spacetime is $n_{I + \bar I} \approx 1 \, \textrm{fm}^{-4}$;
the average instanton size is $\bar\rho \approx 0.3$ fm. The fraction of the 4D volume occupied
by instantons is small, $\kappa \equiv \pi^2 \bar\rho^4 n_{I + \bar I} \approx 0.1$ (packing fraction)
and serves as a parameter for organizing calculations. Approximate descriptions of the QCD vacuum
based on instantons have been developed by abstracting from these features and tested by calculating
correlation functions \cite{Schafer:1996wv,Diakonov:2002fq}. They have also been validated
by comparing with lattice QCD simulations in the cooled or gradient flowed regime
\cite{Chu:1994vi,Bonati:2014tqa,Alexandrou:2017hqw,Athenodorou:2018jwu}.

Chiral symmetry breaking is obtained by studying the propagation of light quarks in the
instanton medium \cite{Diakonov:1985eg}. The zero modes of the individual instantons induce many-fermionic
interactions with a definite spin-flavor structure ('tHooft vertex). In the instanton medium with finite
density a dynamical quark mass $M$ develops, of the order $M^2 \sim \kappa \bar\rho^{-2}$, accompanied
by the formation of a chiral condensate. The effective dynamics can be derived by integrating
over the instanton fields and computing the effective action of the quarks in the $1/N_c$ expansion
(saddle point approximation to the fermionic functional integral,
bosonization) \cite{Nowak:1989at,Diakonov:1995qy,Kacir:1996qn}. 
It is obtained as
\begin{align}
& Z_{\mathrm{eff}} \equiv \int \mathcal{D} U \!\int\mathcal{D}\bar\psi \mathcal{D}\psi \;
\exp\left(i \int d^4 x \, \mathcal{L}_{\mathrm{eff}}\right),
\label{effective_theory}
\\[0ex]
& \mathcal{L}_{\mathrm{eff}} = \bar\psi (x)
[ i\slashed{\partial} - M U^{\gamma_{5}}(x) ] \psi (x),
\\[1ex]
&U^{\gamma_{5}}(x) \equiv \tfrac{1}{2}(1+\gamma_{5}) U(x) + \tfrac{1}{2}(1-\gamma_{5}) U^{\dagger}(x),
\label{U_gamma5_def}
\end{align}
where $\psi, \bar\psi$ are the quark fields
with $N_f$ flavor components and $U(x)$ is the $N_f \times N_f$ matrix-valued
boson field parametrizing the local flavor rotations of the chiral condensate.
Equations~(\ref{effective_theory})--(\ref{U_gamma5_def})
describe the dynamics of the massive quark field modes with
virtualities $p^2 \sim M^2 \ll \bar\rho^{-2}$ (the action is given in Minkowskian form).
The ultraviolet cutoff is realized by the form factors of the instanton zero mode, which suppress
modes with $p^2 \sim \bar\rho^{-2}$ and are omitted here (see references).

The effective action is used to compute correlation functions of light quark operators in the
$1/N_c$ expansion and successfully describes a large body of meson and baryon structure as governed
by chiral symmetry breaking. The pion appears as a Nambu-Goldstone mode with characteristic structure
and interactions; for other applications in the meson sector
see Refs.~\cite{Nowak:1989at,Diakonov:1995qy,Kacir:1996qn}.
The nucleon appears as a mean-field solution with characteristics of a chiral soliton (see following).

\section{Effective operators}
In the instanton vacuum, QCD operators composed from gauge fields sample the classical
gauge fields of the instantons. When integrating over the instantons, such QCD operators are
converted to effective operators in the effective theory after chiral symmetry breaking,
Eq.~(\ref{effective_theory}), expressed in the massive fermion fields and the chiral boson fields
\cite{Diakonov:1995qy,Balla:1997hf,Weiss:2025nzy},
\begin{align}
\mathcal{O}_{\rm QCD}[\bar\psi, \psi, A]
\;\; (\textrm{at scale $\bar\rho^{-1}$})
\rightarrow
\mathcal{O}_{\rm eff}[\bar\psi, \psi, U].
\label{effop}
\end{align}
The effective operators describe the dynamical effects of instanton-induced QCD gauge interactions
in terms of the effective degrees of freedom after chiral symmetry breaking. 

The effective operators representing the QCD quark EMT Eq.~(\ref{emt_f_qcd}) have been derived
in Ref.~\cite{Kim:2023pll}. Their form depends on the spin projection (twist) of the tensor
and the flavor combination of the quark fields. In the following we consider $N_f = 2$
and define flavor-singlet and nonsinglet QCD operators as
\begin{align}
T^{\mu\nu}(x) \equiv \bar{\psi}(x) \gamma^{\mu} i \overleftrightarrow{\nabla}^{\nu} \tau \psi (x),
\end{align}
where $\tau = (1, \tau^3)$ for the combination $u \pm d$.
For the symmetric traceless (twist-2) part of the tensor, the effective operator is
\begin{align}
&\tfrac{1}{2} T^{\{\mu\nu\}}(x) - \tfrac{1}{4} g^{\mu\nu} T^{\rho}_{\;\;\rho}(x) \cr
&=\bar{\psi}(x) \left(  \tfrac{1}{2} \gamma^{ \{ \mu} i \overleftrightarrow{\partial}^{\nu\}}
  - \tfrac{1}{4} g^{\mu\nu} \gamma^\rho i \overleftrightarrow{\partial_\rho} \right) \tau \psi (x).
\label{eff_twist2}
\end{align}
In this part there is no dynamical effect of the instanton field in leading
order of the packing fraction. The effective operator involves only the momentum
(derivative) of the massive quark field in the effective theory.
It coincides with the twist-2 part of the EMT derived from the
effective action Eqs.~(\ref{effective_theory})--(\ref{U_gamma5_def}) using Noether's theorem.
For the antisymmetric (twist-3) part of the tensor, the effective operator is
\begin{align}
\tfrac{1}{2} T^{ [\mu \nu ]}(x) &\equiv \tfrac{1}{2} T^{[\mu \nu]}[\mathrm{kin}]
+ \tfrac{1}{2} T^{[\mu \nu]}[\mathrm{pot}]
\nonumber \\
&= \bar{\psi}(x) \left( 
\tfrac{1}{2} \gamma^{[\mu} i\overleftrightarrow{\partial}^{\nu]}\tau
+ \tfrac{i}{4} M
\sigma^{\mu \nu}   [\tau,U^{\gamma_{5}}(x)] \right) \psi(x),
\label{eff_twist3}
\end{align}
where $[\tau,U^{\gamma_{5}}] \equiv \tau U^{\gamma_{5}} - U^{\gamma_{5}} \tau$.
The first term in Eq.~(\ref{eff_twist3}) involves the momentum of the quark field.
The second term results from the gauge potential in the QCD covariant derivative in Eq.~(\ref{emt_f_qcd}),
which samples the instanton field and gets converted to a chiral spin-flavor
interaction \cite{Kim:2023pll}. We refer to them  as the ``kinetic'' and ``potential''
terms of the effective operator.

The potential term of the twist-3 EMT resulting from instantons has important dynamical effects. 
One immediate effect is that the effective operator Eq.~(\ref{eff_twist3})
obeys the same equations-of-motion relation as the original QCD operator, Eq.~(\ref{eom_qcd}),
\begin{align}
\tfrac{1}{2} T^{ [\mu \nu ]}(x) = -\tfrac{1}{4} \epsilon^{\mu \nu \alpha \beta}
\partial_{\alpha} [\bar{\psi} (x) \gamma_{\beta} \gamma_{5} \tau \psi (x)].
\label{eom_eff}
\end{align}
The relation is now obtained using the effective operator and the equations of motion
of the massive quark field in the effective theory derived from Eqs.~(\ref{effective_theory})--(\ref{U_gamma5_def}).
The effect of the potential term on the QCD orbital AM and the spin sum rule
are studied in the following.

The chiral boson field in the effective theory describes scalar-isoscalar and pseudoscalar-isovector
modes and can be written as ($a = 1,2,3$)
\begin{align}
U(x) = \sigma(x) + i\tau^a \pi^a(x),
\hspace{1em}
\sigma^2 + \textstyle{\sum_a} (\pi^a)^2 = 1.
\end{align}
In the twist-3 effective operator Eq.~(\ref{eff_twist3}) with $\tau = \tau^3$, the potential term
then takes the form
\begin{align}
- \tfrac{i}{2} M \epsilon^{3bc} \pi^b (x) \, \bar{\psi}(x) \sigma^{\mu\nu} \gamma_5 \, \tau^c \psi (x).
\label{eff_twist3_pion}
\end{align}
It describes the interaction of the quark spin with the pseudoscalar-isovector chiral field
and gives rise to a ``chiral magnetic effect'' in the nucleon matrix element (see following).
Note that the coupling is given by the dynamical quark mass, a consequence of chiral symmetry breaking.

\section{Nucleon matrix elements}
The nucleon appears as a mean-field solution of the effective dynamics
Eqs.~(\ref{effective_theory})--(\ref{U_gamma5_def})
in the large-$N_c$ limit \cite{Diakonov:1987ty}. It is characterized by a classical chiral field, which is
time-independent and has a particular spatial form in which the isospin direction is aligned
with the spatial direction (``hedgehog''),
\begin{align}
& U_{\rm cl}(\bm{x}) = \exp[i \bm{n} \cdot \bm{\tau} P(r)]
= \sigma_{\rm cl}(\bm{x}) + i \tau^a \pi_{\rm cl}^a(\bm{x}),
\\[1ex]
& \sigma_{\rm cl}(\bm{x}) = \cos P(r),
\hspace{1em}
\pi_{\rm cl}^a(\bm{x}) = n^a \sin P(r),
\label{hedgehog}
\end{align}
where $\bm{n} \equiv \bm{x}/|\bm{x}|$ is the radial unit vector, $r \equiv |\bm{x}|$, and $P(r)$ is a radial profile
function with $P(0) = \pi$ and $P(\infty) = 0$.
The quarks move in single-particle orbits in the classical chiral field, with the wave functions
$\Phi_{n}(\bm{x})$ and energy levels $E_{n}$ obtained from the Hamiltonian
\begin{align}
& H \equiv - \gamma^0 \gamma^k i \partial_{k} + \gamma^0 M U^{\gamma_{5}}_{\rm cl} (\bm{x}),
\label{hamiltonian}
\\
& H \Phi_{n} (\bm{x}) = E_{n} \Phi_{n} (\bm{x}).
\label{spectrum}
\end{align}
The spectrum includes a discrete level with energy $E_{\rm lev} < M$, and the continuous
spectra of levels with negative energies $E < -M$ and positive energies $E > M$. In the state
with baryon number $B = 1$, the discrete level and the and negative continuum are occupied.
Baryon spin-flavor states are obtained by quantizing the collective (iso-) rotations of the
mean-field state.

The matrix elements of effective operators between nucleon states are computed in the $1/N_c$ expansion,
using the techniques summarized in Ref.~\cite{Christov:1995vm}. The results can be expressed as sums
over the quark single-particle states in the chiral field. This field-theoretical description
of the nucleon preserves completeness and realizes the partonic sum rules, with essential contributions
from the negative energy ``sea'' quarks \cite{Diakonov:1996sr,Diakonov:1997vc}.

We now compute the nucleon matrix elements of the QCD AM operators Eqs.~(\ref{L_qcd}),
(\ref{S_qcd}) and (\ref{J_qcd}), describing the orbital, spin and total quark AM.
The effective operators in the chiral theory are obtained by replacing the QCD quark EMT
with the corresponding effective operators Eqs.~(\ref{eff_twist2}) and (\ref{eff_twist3}).
The flavor-nonsinglet nucleon matrix elements appear in leading order in the $1/N_c$ expansion,
\begin{align}
L_{u-d}, \, S_{u-d}, \, J_{u-d} = \mathcal{O}(N_{c}).
\end{align}
The nucleon matrix elements can be expressed as sums of diagonal matrix elements of the effective
operators in quark single-particle states. As such they permit a simple interpretation in terms
of first-quantized quark single-particle operators.

The total AM, Eq.~(\ref{J_qcd}), is computed with the twist-2 EMT, Eq.~(\ref{eff_twist2}),
in which there is no chiral interaction effect. We obtain the nucleon matrix element as
\begin{align}
J_{u-d} &= \frac{N_{c}}{3} \int d^{3}x 
\sum_{n=\mathrm{occ}}
\Phi_{n}^\dagger  (\bm{x}) \left[ -\tfrac{1}{2} L^{3} \tau^3
\right.
\nonumber \\
& \left. \hspace{2em} - \tfrac{1}{2} E_n (\bm{x}
\times \bm{\Sigma})^{3} \gamma_{5} \tau^3 \right] \Phi_{n} (\bm{x}).
\label{J_nucleon}
\end{align}
The summation runs over all occupied quark levels, including the discrete level and the
negative-energy sea.
In the expression in brackets, $\bm{L} \equiv \bm{x} \times \bm{p}$ is the single-particle
AM operator and $\Sigma^j \equiv -\gamma^0 \gamma^j \gamma_5 = i \sigma^{0j} \gamma_5$
is the spin operator. (The sign in front of the single-particle operators is due to
the particular spin-isospin coupling in the single-particle wave functions; see following.)
The spin AM, Eq.~(\ref{S_qcd}), is computed with the antisymmetric twist-3 EMT,
Eq.~(\ref{eff_twist2}); because of the equations-of-motion relation Eq.~(\ref{eom_eff})
it is equal to the matrix element of the axial current operator in the effective theory.
We obtain
\begin{align}
&S_{u-d} = \frac{N_{c}}{3}  \int d^{3}x \sum_{n=\mathrm{occ}} \Phi_{n}^\dagger  (\bm{x}) 
\left[ -\tfrac{1}{2} \Sigma^{3} \tau^{3} \right] \Phi_{n}  (\bm{x}).
\label{S_nucleon}
\end{align}
The orbital AM is the most interesting. It is computed with the symmetric (twist-2) and
antisymmetric (twist-3) parts of the EMT; the effective operator of the latter contains the
instanton-induced potential term, Eq.~(\ref{eff_twist3}). We obtain
\begin{align}
& L_{u-d} = L_{u-d}[\mathrm{kin}] + L_{u-d}[\mathrm{pot}],
\label{L_nucleon}
\\[1ex]
& L_{u-d}[\mathrm{kin}] = \frac{N_{c}}{3}  \int d^{3}x \sum_{n=\mathrm{occ}}
\Phi_{n}^\dagger  (\bm{x}) [ -L^{3} \tau^{3} ] \Phi_{n} (\bm{x}),
\label{L_kin}
\\
& L_{u-d}[\mathrm{pot}] =  \frac{N_{c} M}{6}  \int d^{3}x \,
\epsilon^{3jk} \epsilon^{3bc} x^j \pi_{\rm cl}^b(\bm{x})
\nonumber \\
& \hspace{4em} \times \sum_{n=\mathrm{occ}} \Phi_{n}^\dagger  (\bm{x})
\left[ \gamma^{0} \Sigma^k \tau^c \right]
\Phi_{n} (\bm{x}).
\label{L_pot}
\end{align}
Equation~(\ref{L_kin}) results from the kinetic term of the twist-3 effective operator Eq.~(\ref{eff_twist3}).
Equation~(\ref{L_pot}) results from the potential term and describes the interaction of the quarks
with the classical chiral field, see Eq.~(\ref{eff_twist3_pion}).
With the specific form of the field as in Eq.~(\ref{hedgehog}), the expression in the
integrand becomes
\begin{align}
\epsilon^{3jk} \epsilon^{3bc} x^j \pi_{\rm cl}^b(\bm{x})
\rightarrow
r \sin{P(r)} \, \epsilon^{3jk} \epsilon^{3bc} n^j n^b.
\end{align}
The interpretation of this effect is developed in the following.

The nucleon matrix elements of the total, spin, and orbital AM
in the effective theory satisfy the spin sum rule Eq.~(\ref{sumrule_qcd}),
\begin{align}
J_{u - d} = S_{u - d} + L_{u - d}.
\end{align}
This can be shown by combining the single-particle operators in Eqs.~(\ref{S_nucleon}) and (\ref{L_nucleon})
and using the equations of motion in the effective theory in the form of the eigenvalue equation
Eq.~(\ref{spectrum}). The instanton-induced potential term in the twist-3 operator is essential
for bringing about this result. The sum rule attests to the consistency of the approximations
in our framework.

Our findings resolve the problem noted in the earlier study of Ref.~\cite{Wakamatsu:2005vk}, where only
the kinetic part of the effective twist-3 operator was included and a violation of the spin sum rule
was observed. The inclusion of the potential term of the twist-3 operator,
which represents QCD interaction effects at the level of the effective theory, naturally restores
the spin sum rule.

\begin{table}[t]
\centering
\setlength{\tabcolsep}{8pt}
\renewcommand{\arraystretch}{1.2}
\begin{tabular}{l | r r | r | } 
\hline
\hline
 & lev  & sea & total  \\
\hline
$J_{u-d}$  & $0.248$ & $-0.008$ & $0.240$
\\[-.5ex]
$S_{u-d}$  & $0.362$ & $0.130$ & $0.493$ 
\\[-.5ex]
$L_{u-d}$  & $-0.114$ & $-0.139$ & $-0.252$ 
\\[-.5ex]
$L_{u-d}[\mathrm{kin}]$  & $0.138$ & $-0.182$ & $-0.044$ 
\\[-.5ex]
$ L_{u-d}[\mathrm{pot}]$  & $-0.252$ & $0.044$ & $-0.208$
\\ 
\hline 
\hline
\end{tabular}
\caption{Numerical results for the flavor-nonsinglet AM in the proton in the
large-$N_c$ mean field picture, as given by Eqs.~(\ref{J_nucleon}), (\ref{S_nucleon}),
and (\ref{L_nucleon})--(\ref{L_pot}).
For reference we quote the contributions of
the discrete level (lev), the negative continuum (sea), and the total result.
\label{tab:num}
}
\end{table}

We now evaluate numerically the total, spin, and orbital AM given by
Eqs.~(\ref{J_nucleon}), (\ref{S_nucleon}) and (\ref{L_nucleon}) and study
the AM decomposition. For the orbital AM we separately compute
the contributions of the kinetic and potential term in Eq.~(\ref{L_nucleon}).
The results are summarized in Table~\ref{tab:num}.
One observes:
(i)~The dominant contribution to $L_{u - d}$ comes from the
potential term, showing the importance of the chiral interaction effect.
(ii)~A large negative $L_{u - d}$ is obtained overall.
(iii)~The large negative $L_{u - d}$ partly cancels the large positive $S_{u - d}$,
resulting in a reduced positive value of $J_{u - d}$.
(For the numerical evaluation we use $M =$ 350 MeV, the Pauli-Villars ultraviolet
cutoff of Ref.~\cite{Diakonov:1997vc}, and the soliton profile $P(r)$
determined by self-consistent minimization of the static energy of the baryon with these parameters.)

We can compare our results with recent lattice QCD calculations.
EMTC \cite{Alexandrou:2020sml} reports $J_{u-d} = 0.161(24)$, $S_{u-d} = 0.644(12)$,
corresponding to $L_{u-d} = -0.483$ (scale $\mu =$ 2 GeV).
$\chi$QCD \cite{Wang:2021vqy} reports $J_{u-d} = 0.151$, $S_{u-d} = 0.672$,
corresponding to $L_{u-d} = -0.476$ (same scale). 
Our results in Table~\ref{tab:num} are consistent with the lattice values,
given the expected accuracy of our approximations (effective dynamics,
leading-order $1/N_c$ expansion). The normalization scale of our results is
roughly the inverse instanton size $\bar\rho^{-1} \approx$ 600 MeV; scale evolution
is moderate in the flavor-nonsinglet sector \cite{Thomas:2008ga}, and we neglect it in the comparison here.
The main point is that our results explain the large $L_{u-d} < 0$ observed in the lattice calculations.
Note that the flavor-nonsinglet spin $S_{u - d} = \tfrac{1}{2} g_A$ is known to receive
substantial $1/N_c$ corrections \cite{Wakamatsu:1993nq,Christov:1993ny},
which can be expected to affect also $L_{u-d}$ and $J_{u-d}$.

\section{Chiral magnetic effect}
The potential term of the flavor-nonsinglet quark orbital AM in the effective theory permits
an interesting interpretation as a ``chiral magnetic effect.''
The expression Eq.~(\ref{L_pot}) involves the single-particle matrix elements of the quark spin operator.
The sum with $\epsilon^{3jk}$ (cross product) samples the transverse spin components $k = 1, 2$.
The interaction with the chiral mean field converts the quark transverse spin
into a contribution to the longitudinal quark AM (3-direction). 
This effect is similar to spin precession in an external magnetic field.

\begin{figure}[t]
\centering
\includegraphics[width=0.6\columnwidth]{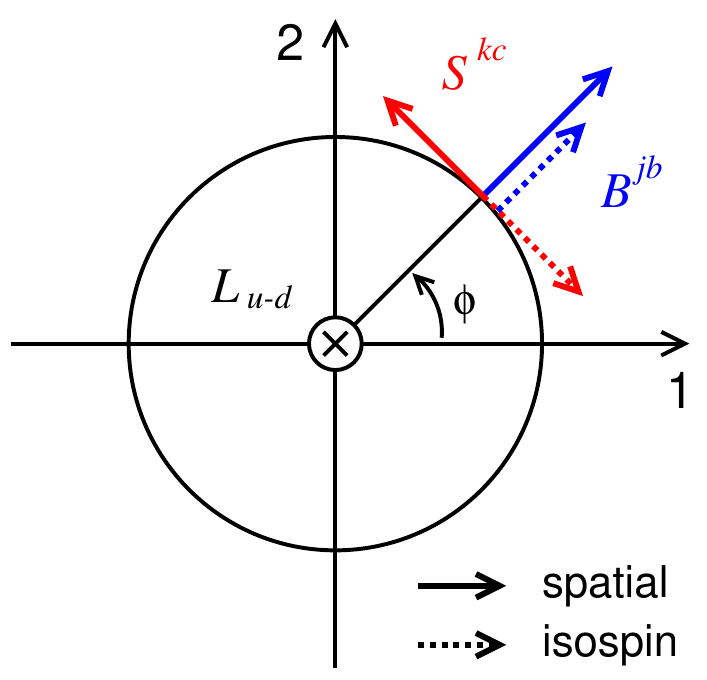}
\caption{Visualization of the chiral magnetic effect generating the flavor-nonsinglet orbital angular
momentum in the large-$N_c$ proton, Eq.~(\ref{L_chiralmagnetic}). Shown is the transverse $1, 2$ plane,
with the 3-axis pointing toward the observer. Spatial and spin vectors are indicated by solid lines,
isospin vectors by dashed lines. \textit{Red arrows:} Tangential components of quark spin and isospin.
\textit{Blue arrows:} Radial components of magnetic field in space and isospin.}
\label{fig:chiralmagnetic}
\end{figure}

To exhibit the analogy, we write Eq.~(\ref{L_pot}) in the form
\begin{align}
L_{u-d}[\mathrm{pot}] &=   \int d^{3}x \;
\epsilon^{3jk} \epsilon^{3bc} B^{jb}(\bm{x}) \, S^{kc}(\bm{x}),
\label{L_chiralmagnetic}
\\
B^{jb}(\bm{x}) &=  M x^j \pi_{\rm cl}^b (\bm{x}) = M r \sin P(r) \, n^j n^b,
\label{B}
\\
S^{kc}(\bm{x}) &\equiv  \frac{N_{c}}{3} \sum_{n=\mathrm{occ}}
\Phi_{n}^\dagger  (\bm{x})
\left[ \tfrac{1}{2} \gamma^{0} \Sigma^k \tau^c \right]
\Phi_{n} (\bm{x}).
\label{spin_current}
\end{align}
$B^{jb}$ is the ``chiral magnetic field.'' The vector index is carried by the
coordinate vector appearing in the original definition of the QCD AM,
Eq.~(\ref{L_qcd}); the parity and isospin are fixed by the classical pion field.
$S^{kc}$ is the spin-isospin current of the quarks in the mean-field state.
Because of the cross products in Eq.~(\ref{L_chiralmagnetic}), and because the
spin and isospin of the magnetic field Eq.~(\ref{B}) are along the radial direction,
Eq.~(\ref{L_chiralmagnetic}) samples the transverse tangential components of the
quark current in spin and isospin, i.e., the components along
\begin{align}
\bm{e}_\phi \equiv -\sin\phi \, \bm{e}_1 + \cos\phi \, \bm{e}_2,
\hspace{1em}
\bm{e}_\phi \cdot \bm{n} = 0.
\end{align}
The radial chiral magnetic field converts the tangential quark spin and isospin into a
longitudinal isovector angular momentum (see Figure~\ref{fig:chiralmagnetic}).
The conversion happens simultaneously in the spatial and isospin vectors,
a remarkable consequence of the hedgehog symmetry of the large-$N_c$ mean field solution.
%
%

The sign of the potential AM, Eq.~(\ref{L_chiralmagnetic}), is determined by whether the tangential
quark spin and isospin components are on average aligned or antialigned. To investigate this,
we can inspect the contribution of the discrete single-particle level. Its bispinor wave function is
given by
\begin{align}
\Phi_{\rm lev} (\bm{x}) &= \frac{1}{\sqrt{4\pi}}
\left(
\begin{array}{r} f_0 (r) \\[0ex] 
\displaystyle
-i \bm{n}\cdot\bm{\sigma} \, f_1 (r) 
\end{array} \right) \chi ,
\label{level_spinor}
\end{align}
where $\chi$ is the spinor-isospinor with spin $\tfrac{1}{2}$ and isospin $\tfrac{1}{2}$ coupled to
zero total, $(\bm{\sigma} + \bm{\tau}) \chi = 0$, with $\chi^\dagger \chi = 1$, and
$f_0, f_1$ are the radial wave functions \cite{Diakonov:1987ty,Kim:2024cbq}.
For this level the transverse tangential spin-isospin current is
\begin{align}
e_\phi^k \, e_\phi^c \, S^{kc}_{\rm lev}
&= e_\phi^k \, e_\phi^c \, \frac{N_c}{3} \, \Phi_{\rm lev}^\dagger (\bm{x}) 
\left[ \tfrac{1}{2} \gamma^{0} \Sigma^k \tau^c \right]
\Phi_{\rm lev} (\bm{x})
\nonumber \\
&= - \frac{N_c}{24\pi} [f_0^2 + f_1^2] (r),
\label{spin_current_level}
\end{align}
which is manifestly negative. It shows that the tangential quark spin and isospin components
are antialigned, so that $L_{u-d}[\textrm{pot}]$ points in the negative 3-direction
(see Figure~\ref{fig:chiralmagnetic}). Equation~(\ref{spin_current_level}) also shows
that the tangential spin-isospin current is quadratic in the large bispinor component $\propto f_0$
and not a small relativistic effect. The picture thus naturally explains the negative
sign and large magnitude of $L_{u-d}$.

The picture proposed here exhibits a formal analogy with the chiral magnetic effect in hot
or dense QCD matter \cite{Kharzeev:2004ey,Fukushima:2008xe}. A chiral field excited in
high-energy collisions implies a local violation of parity invariance and manifests itself
as an asymmetry of the observed charged pion distribution. In our case the chiral field is the
one naturally present in large-$N_c$ nucleon. [It is interesting to note that the isospin
components at work in Eq.~(\ref{L_chiralmagnetic}) are the $1, 2$ components
describing the charged pion modes.]

\section{Bosonized representation}
The nucleon matrix elements obtained from the mean-field solution show a large flavor-nonsinglet
orbital AM, which partly cancels the spin and reduces the value of the total AM.
This pattern can be explained by general properties of the chiral dynamics and is model-independent.
To show this, we convert the effective dynamics to bosonized form and perform a derivative expansion
of the AM operators (chiral soliton picture). 

Integrating over the fermion fields in Eq.~(\ref{effective_theory}), the effective dynamics is
converted to bosonized form
\begin{align}
& Z_{\mathrm{eff}} = \int \mathcal{D} U \, \exp (i W_{\mathrm{eff}}[U]) ,
\label{effective_theory_bosonized}
\\
& W_{\mathrm{eff}}[U] \equiv \mathrm{log \, Det}[i \slashed{\partial} - M U^{\gamma_{5}}]
\nonumber \\
& = \int d^{4} x \left\{ \tfrac{1}{4} F_\pi^2
\operatorname{tr} [\partial^{\alpha} U^{\dagger} \partial_{\alpha} U]
+ \textrm{terms $\mathcal{O}(p^4, p^6, ...)$} \right\}.
\label{chiral_lagrangian}
\end{align}
The effective action is given by the fermion determinant in the background of the chiral field.
Expansion in derivatives of the chiral field reproduces the chiral Lagrangian;
the form of the terms is dictated by chiral symmetry \cite{Gasser:1983yg}; the dynamical constants
such as $F_\pi$ are obtained from the instanton vacuum parameters ($M, \bar\rho^{-1}$) \cite{Diakonov:1987ty}.
By the same procedure, the instanton-induced effective operators Eq.~(\ref{effop}) are converted
to bosonized operators,
\begin{align}
\mathcal{O}_{\rm eff}[\bar\psi, \psi, U]
\rightarrow
\mathcal{O}_{\rm eff, bosonized}[U],
\end{align}
which can be expanded in derivatives of the chiral fields; see Refs.~\cite{Kim:2023pll,Kim:2024cbq,Balla:1997hf}
for details. In this way the QCD quark EMT and the AM operators can be converted to bosonized chiral operators
and analyzed using ``chiral counting'' arguments.

For the flavor-nonsinglet twist-2 and 3 operators, Eqs.~(\ref{eff_twist2}) and (\ref{eff_twist3}),
we obtain at $\mathcal{O}(p^2)$
\begin{align}
\tfrac{1}{2} T_{u-d}^{\{\mu \nu\}} - \textrm{trace} = 0,
\hspace{1em}
\tfrac{1}{2} T^{[\mu \nu]}_{u - d} = -\frac{F_\pi^2}{8i} \epsilon^{\mu\nu\alpha\beta} \partial_{\alpha}
\operatorname{tr}[\tau^{3}(R_{\beta} - L_{\beta})],
\label{bosonized_nonsinglet}
\end{align}
where $R_{\mu}, L_{\mu} \equiv U \partial_{\mu} U^{\dagger}, \, U^{\dagger} \partial_{\mu} U$.
The twist-2 part (symmetric tensor)
is zero at $\mathcal{O}(p^2)$. The twist-3 part (antisymmetric tensor) is obtained
as the derivative of the isovector axial current of the chiral field, so that the equations-of-motion
relation Eq.~(\ref{eom_eff}) is satisfied by the chiral operators at $\mathcal{O}(p^2)$.
While this is as expected, it is obtained from the actual bosonization of the twist-3
effective quark operator Eq.~(\ref{eff_twist3}), a very gratifying result.

The nucleon in the bosonized theory is described by the classical chiral field of Eq.~(\ref{hedgehog})
(chiral soliton) \cite{Zahed:1986qz}.
The matrix elements of the chiral operators Eq.~(\ref{bosonized_nonsinglet})
are computed by evaluating the operators in the classical chiral field and quantizing the collective
(iso-) rotations. The fact that $\tfrac{1}{2} T_{u-d}^{\{0k\}} = 0$
at $\mathcal{O}(p^2)$ means that $J_{u-d} = 0$. At the same time, it implies that $T_{u-d}^{0k}
= \tfrac{1}{2} T_{u-d}^{[0k]}$, and therefore $L_{u-d} = -S_{u-d}$. Altogether, the chiral structure of
Eq.~(\ref{bosonized_nonsinglet}) implies
\begin{align}
J_{u-d} = 0, \quad S_{u-d}=-L_{u-d} \quad \mathrm{at} \quad \mathcal{O}(p^{2}).
\label{chiral_nonsinglet}
\end{align}
It explains the cancellation of spin and orbital AM in the flavor non-singlet sector.
The matrix elements in leading order of $1/N_c$ are obtained as
\begin{align}
S_{u-d} = -L_{u-d} = \frac{F_\pi^2}{9} \int d^{3}r \left[ -P'(r) - \frac{ \sin{2P(r)}}{r}\right],
\end{align}
with $F_\pi^2 = \mathcal{O}(N_c)$. Numerical evaluation gives $S_{u-d} = -L_{u-d} \approx 0.13$,
which is qualitatively consistent with the pattern of the numerical results in Table~\ref{tab:num}.
The $\mathcal{O}(p^{2})$ bosonized calculation
is intended to explain the results of the full mean-field calculation, not as a realistic numerical estimate.

It is interesting to contrast this with the flavor-singlet sector.
Here the symmetric tensor operator is $\mathcal{O}(p^2)$ and of the same form as the EMT
derived from the chiral Lagrangian Eq.~(\ref{chiral_lagrangian}) using Noether's theorem.
The antisymmetric tensor operator is zero at this level. At $\mathcal{O}(p^2)$ we obtain 
\begin{align}
&\tfrac{1}{2} T^{ \{\mu \nu\} }_{u + d} - \mathrm{trace}
=-\frac{F_\pi^{2}}{2} \operatorname{tr}[L^{ \mu}L^{\nu }
- \tfrac{1}{4} L^{\beta}L_{\beta} g^{\mu\nu}],
\hspace{1em}
\tfrac{1}{2} T^{ [\mu \nu] }_{u + d} = 0.
\label{bosonized_singlet}
\end{align}
It implies that $J_{u + d}$ now appears at $\mathcal{O}(p^2)$.
It also implies that $T^{0k}_{u + d} = \tfrac{1}{2} T^{\{ 0k \}}_{u + d}$ at order $\mathcal{O}(p^2)$,
so that $L_{u + d}$ also appears at $\mathcal{O}(p^2)$ and satisfies
$L_{u + d} = J_{u + d}$. The flavor-singlet axial current is of higher order in the
chiral expansion, so that $S_{u + d} = 0$ at $\mathcal{O}(p^2)$.
Altogether, the chiral structure now implies
\begin{align}
&J_{u+d}=L_{u+d}, \quad S_{u+d}=0 \quad \mathrm{at} \quad \mathcal{O}(p^{2}).
\end{align} 
The total AM comes entirely from the orbital AM of the quarks; the spin zero at this level
of the chiral expansion. Our results reproduce the traditional argument for a small value
of the flavor-singlet quark spin based on the soliton picture of the nucleon,
put forward during the ``proton spin crisis'' \cite{Brodsky:1988ip}.

In the flavor-singlet channel one also needs to consider the gluon contribution to the QCD AM.
Chiral counting cannot discriminate between $J_{u+d}$ and $J_g$; both are of the same order
in the chiral expansion, and the splitting between the two is a dynamical question.
In the instanton vacuum the twist-2 gluon operator is suppressed relative to the twist-2
quark operator by a power of the instanton packing fraction, $J_g = \mathcal{O}(\kappa),
J_{u+d} = \mathcal{O}(1)$. In this context it is justified to identify the twist-2 flavor-singlet
quark EMT with the total EMT as done in Eq.~(\ref{bosonized_singlet}). It is validated by the
fact that the bosonization of the twist-2 flavor-singlet quark EMT Eq.~(\ref{eff_twist2})
indeed reproduces Eq.~(\ref{bosonized_singlet}), and that the nucleon matrix element computed
with this operator satisfies the AM sum rule $J_{u+d} = \tfrac{1}{2}$.

\section{Summary}
The results of this study can be summarized as follows:
(i)~Instanton-induced gauge interactions
have a large effect on the flavor-nonsinglet twist-3 quark EMT. At the level of the effective theory
resulting from chiral symmetry breaking, the effect is represented by chiral interactions
(potential term) in the effective twist-3 tensor operator.
(ii)~The chiral interactions in the twist-3 quark EMT give rise to a large
flavor-nonsinglet orbital AM $L_{u - d}$.
In the large-$N_c$ limit, the effect takes the form of an interaction
of the quarks with the classical chiral field in the nucleon and permits
an interpretation as a chiral magnetic effect analogous to that in hot/dense matter.
(iii)~The $L_{u - d}$ arising in this way is negative and partly cancels the large positive
spin $S_{u - d}$. This explains reduced value of total $J_{u - d}$ observed in lattice
QCD calculations. 
(iv)~In the bosonized low-energy theory, the observed pattern of spin, orbital and total AM is explained
by the chiral counting of the bosonized effective operators, in the flavor-nonsinglet and
singlet sectors.

Altogether, the results show a close connection between flavor-nonsinglet quark AM and
chiral symmetry breaking in QCD. Theoretical and experimental studies of the flavor-nonsinglet
quark AM can provide essential insight into the effective low-energy dynamics and should be
pursued with high priority.

\section*{Acknowledgments}
J.-Y.~Kim acknowledges support from Inha University Research Grant 2026, No.~77302-1.
The work of H.-Y.~W.\ is supported by the France
Excellence scholarship through Campus France funded
by the French government (Minist\`ere de l'\ Europe et des
Affaires \'Etrang\`eres), Grant No. 141295X.

This material is based upon work supported by the U.S.~Department of
Energy, Office of Science, Office of Nuclear Physics under contract
DE-AC05-06OR23177. The research reported here is connected with
the Topical Collaboration ``3D quark-gluon structure of hadrons:
mass, spin, tomography'' (Quark-Gluon Tomography Collaboration),
supported by the U.S.~Department of Energy, Office of Science,
Office of Nuclear Physics under contract DE-SC0023646.

\bibliography{am_decomposition}

\end{document}